\documentclass[prl,aps,twocolumn, superscriptaddress]{revtex4}
\usepackage{amsmath}
\usepackage{amssymb}
\usepackage[english]{babel}
\usepackage{graphicx}

\makeatletter
\let\NAT@bibsetnum\NATx@bibsetnum
\renewenvironment{thebibliography}[1]{
 \bibsection\parindent \z@\bibpreamble\bibfont\list
   {\@biblabel{\arabic{NAT@ctr}}}{\@bibsetup{#1}}
}{
 \NAT@endthebibliography
}
\makeatother

\begin{document}

\title{Majorana state on the surface of a disordered 3D topological insulator}

\author{P.\ A.\ Ioselevich}
\affiliation{L.\ D.\ Landau Institute for Theoretical Physics, Kosygin str.\ 2,
Moscow, 119334 Russia}
\affiliation{Moscow Institute of Physics and Technology, Institutsky per.\ 9,
Dolgoprudny, 141700 Russia}

\author{P.\ M.\ Ostrovsky}
\affiliation{Max Planck Institute for Solid State Research, Heisenbergstr.\ 1,
70569 Stuttgart, Germany}
\affiliation{L.\ D.\ Landau Institute for Theoretical Physics, Kosygin str.\ 2,
Moscow, 119334 Russia}

\author{M.\ V.\ Feigel'man} 
\affiliation{L.\ D.\ Landau Institute for Theoretical Physics, Kosygin str.\ 2,
Moscow, 119334 Russia}
\affiliation{Moscow Institute of Physics and Technology, Institutsky per.\ 9,
Dolgoprudny, 141700 Russia}

\begin{abstract}
We study low-lying electron levels in an ``antidot'' capturing  a coreless
vortex on the surface of a three-dimensional topological insulator in the
presence of disorder. The surface is covered with a superconductor film  with a
hole of size $R$ larger than coherence length, which induces superconductivity
via proximity effect. Spectrum of electron states inside the hole is sensitive
to disorder, however, topological properties of the system give rise to a robust
Majorana bound state at zero energy. We calculate the subgap density of states
with both energy and spatial resolution using the supersymmetric sigma model
method. Tunneling into the hole region is sensitive to the Majorana level and
exhibits resonant Andreev reflection at zero energy.
\end{abstract}

\maketitle

Topological insulators and superconductors are very peculiar materials with a
gap in the bulk electron spectrum and a low-lying branch of subgap excitations
on their surface (see \cite{reviewQiZhang} for a review). This surface metallic
state appears due to topological reasons and is robust with respect to any
(sufficiently small) perturbations. In particular, topological properties
prevent these surface states from Anderson localization. One common example of a
topological insulator is a two-dimensional system in the integer quantum Hall
effect regime. The bulk of such a system has a spectral gap between successive
Landau levels and is hence an insulator. At the same time quantized Hall
conductance appears due to a fixed integer number of chiral propagating edge
modes on the background of the bulk gap.  This type of materials are referred to
as $\mathbb{Z}$ topological insulators.

Another type of topological insulator is realized in three-dimensional (3D)
semiconductor crystals with sufficiently strong spin-orbit interaction (BiSb,
BiTe, BiSe, strained HgTe etc). The spin-orbit interaction leads to inversion of
the spectral gap. As a result subgap surface excitations appear with a
dispersion of the massless Dirac type. The topological invariant in these
materials has a $\mathbb{Z}_2$ nature. When the number of surface states is
odd, one of them always remains gapless due to topological protection.

A general classification of topological insulators was developed in Refs.\
\cite{Kitaev, Schnyder} based on the symmetry of the underlying Hamiltonian. The
quantum Hall effect is an example of a topologically nontrivial state of the
unitary symmetry (class A of the Altland-Zirnbauer classification
\cite{AltlandZirnbauer}) in 2D. Strong spin-orbit interaction leads to a
topological state in the symplectic class (AII) in 3D. Another important example
is the 1D topological superconductor of the class BD symmetry (superconductor
with both time-reversal and spin rotation symmetries broken). The topologically
protected mode in this case is zero-dimensional and is known as the Majorana
bound state (MBS). It appears, in particular, in the core of an Abrikosov vortex
in the spinless $p$-wave superconductor \cite{Ivanov00}. A similar MBS
appears~\cite{FuKane2008} in a vortex in an ordinary $s$-wave superconductor
brought in contact with the surface of a 3D $\mathbb{Z}_2$ topological
insulator, which corresponds to the symmetry class AII. The MBS appears as a
descendant of the topologically protected massless Dirac state on the free
surface of the topological insulator.

There are two general methods to observe the Majorana level. One of them relates
to the anomalous Josephson effect \cite{kane4pi, IoselFeig2011}. Another, and a
more direct, way involves tunneling into the region where the MBS is supposed
to be localized \cite{kouwenhoven}. The differential conductance in such a
tunneling experiment yields the local density of states with spatial and energy
resolution. The Majorana state in the vortex core manifests itself by resonant
Andreev reflection at zero energy \cite{LeeNg}.

\begin{figure}
\includegraphics[width=0.9\columnwidth]{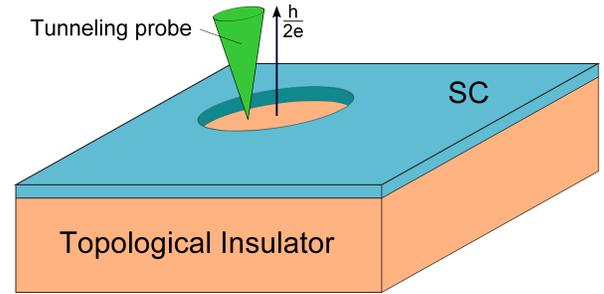}
\caption{(Color online) A sketch of the considered setup. Three-dimensional
topological insulator is covered by a superconducting film with a hole of radius
$R$. The gap in the density of states is induced on the covered surface forming
an antidot and confining surface excitations inside the hole. External magnetic
field induces an Abrikosov vortex inside the hole. The spectrum of energy
eigenstates inside the hole acquires a Majorana zero level which can be
accessed in the tunneling measurement.}
\label{fig_setup}
\end{figure}

In this Letter we study local tunneling conductance for a setup depicted in
Fig.\ \ref{fig_setup}. A superconducting (with a spectral gap $\Delta$)  film 
with a circular hole of the radius $R$ is deposited on the surface of a 3D
topological insulator, e.g., Bi$_2$Te$_3$ crystal. Perpendicular magnetic field
applied to the system produces a vortex pinned to the hole. The radius $R$ is
supposed to be relatively large, $R \geq (l, \xi_0)$, where $l$ is the mean
free path for topological insulator surface states and $\xi_0$ is the
superconducting coherence length of the film. The condition $R \geq \xi_0$
allows one to avoid \cite{Melnikov} the abundance of low-lying Caroli-de
Gennes-Matricon states whose presence in the core of the Abrikosov vortex
complicates observation of the $E=0$  Majorana state. Under these conditions
disorder is relevant for the states localized in the hole region. Although the
Majorana state is protected against disorder, i.e. its energy stays exactly
zero, the spatial distribution of its wavefunction $\psi_0(\mathbf{r})$, and
thus the local tunneling conductance are sensitive to disorder. The aim of this
Letter is to calculate the tunneling conductance in the ``dirty-limit'' with
spatial and energy resolution, and to identify the effects of the Majorana
zero-energy state in the presence of strong disorder.

The problem is described by the following Bogolyubov-de Gennes (BdG)
Hamiltonian in polar coordinates $r$, $\varphi$:
\begin{equation}
 H
  = \begin{pmatrix}
      H_0 & \Delta(r) e^{i \varphi} \\
      \Delta(r) e^{-i \varphi} & - H_0
    \end{pmatrix},
 \quad
 H_0
  = v_0\, \mathbf{s} \cdot \mathbf{p} + V(\mathbf{r}) - \mu.
 \label{H}
\end{equation}
Here the Hamiltonian $H_0$ describes the dynamics of surface excitations in a
topological insulator without a superconducting layer. The Fermi velocity of
surface electrons is denoted by $v_0$, $\mathbf{s}$ is the spin operator, and 
$V(\mathbf{r})$ is the random disorder potential. The Fermi level is shifted
from the Dirac point by $\mu$. The vector potential term in $H_0$ is neglected
due to smallness of magnetic field. We assume the dirty limit with a disorder
induced mean free path $l \ll \xi = \sqrt{\hbar D/\Delta}$ and a relatively
large hole with $R \gg \xi$; here $D = v_0^2 \tau/2$. These conditions allow us
to use a step-like radial dependence of the order parameter:
\begin{equation}
 \Delta(r)
  = \begin{cases}
      0, & r < R, \\
      \Delta, & r > R.
    \end{cases}\label{Delta}
\end{equation}

Inside the hole, the order parameter is zero and electron dynamics is governed
by $H_0$. This Hamiltonian possesses time-reversal symmetry of the symplectic
type, $H_0 = s_y H_0^* s_y$, and hence belongs to the symplectic symmetry class
AII. Proximity to the superconductor induces a gap in the electronic spectrum
outside the hole. This gap effectively confines the low-lying excitations and
imposes boundary conditions for the Hamiltonian $H_0$ at $r = R$. These boundary
conditions break time-reversal symmetry due to the spatially rotating phase of
the order parameter. At the same time the total BdG Hamiltonian $H$ acquires a
specific particle-hole symmetry:
\begin{equation}
 H
  = -s_y \tau_y H^* s_y \tau_y.
 \label{PH}
\end{equation}
Here $\tau_y$ is the Pauli matrix acting in Nambu-Gor'kov (NG) space. Identity
(\ref{PH}) implies a symmetry of the electron spectrum. For each eigenstate
$\psi$ with energy $E$ there is a conjugate eigenstate $s_y \tau_y \psi^*$ with
energy $-E$.

The particle-hole symmetry (\ref{PH}) defines the class BD. On the level of
random matrices there are two distinct versions of this class with even and odd
number of eigenstates referred to as D and B class, respectively. In class B,
one unpaired eigenstate has exactly zero energy and is self-conjugate: $\psi =
s_y \tau_y \psi^*$. The BdG Hamiltonian counts every physical excitation twice
due to the doubling in NG space. An unpaired level is thus ``half'' of a true
excitation --- a Majorana state.

We will calculate the density of states inside the hole with the help of the
supersymmetric non-linear sigma model. Two-dimensional Dirac fermions with
potential disorder are described by a very peculiar model of the class AII with
$\mathbb{Z}_2$ topological term \cite{Ostrovsky}. This topological term appears
as a consequence of the chiral anomaly of Dirac fermions. We will consider the
minimal model operating with the $8 \times 8$ supermatrix $Q$. Apart from
Fermi-Bose superspace, this matrix operates in the space of retarded-advanced
(RA) fields and in a specific time-reversal (TR) space introduced to take into
account Cooperons and diffusons on equal footing. RA space is completely
analogous to the NG space (see below); we denote Pauli matrices in this space by
$\tau$. Notation $\sigma$ is used for Pauli matrices in TR space. A detailed
derivation of the sigma model can be found in \cite{supplement}.

In class AII, the matrix $Q$ obeys the non-linear constraint $Q^2 = 1$ and the
linear constraint
\begin{equation}
 Q
  = \bar Q
  \equiv C Q^T C^T,
 \qquad
 C
  = \tau_x \begin{pmatrix}
      \sigma_x & 0\\
      0 & i \sigma_y
    \end{pmatrix}_\text{FB}.
 \label{barQ}
\end{equation}
As a result, $Q$ contains $8$ commuting and $8$ anticommuting (Grassmann)
variables in total. Commuting degrees of freedom parameterize diagonal
fermion-fermion (F) and boson-boson (B) blocks of $Q$. The general $Q$-matrix
can be decomposed as 
\begin{equation}
 Q
  = U^{-1} \begin{pmatrix}
      Q_\text{F} & 0 \\
      0 & Q_\text{B}
    \end{pmatrix} U,
 \label{UQU}
\end{equation}
where the central part contains only commuting variables while Grassmann
parameters define the unitary supermatrix $U$. We will use the following
explicit form of the central part of $Q$ in terms of eight angle parameters:
\begin{align}
 Q_\text{F}
  &= [ \tau_z \cos \theta_f + \sigma_z \sin \theta_f (
       \tau_x \cos \phi_f + \tau_y \sin \phi_f
     )] \notag\\
  &\;\;\,\times [ \sigma_z \cos k_f + \tau_z \sin k_f (
       \sigma_x \cos \chi_f + \sigma_y \sin \chi_f
     )], \label{qf} \\
 Q_\text{B}
  &= \tau_z \cos \theta_b [\sigma_z \cos k_b + \sin k_b (
       \sigma_x \cos \chi_b + \sigma_y \sin \chi_b
     )] \notag\\
  &\;\;\,+ \sin \theta_b (\tau_x \cos \phi_b + \tau_y \sin \phi_b ). \label{qb}
\end{align}
This representation fulfills all the constraints imposed by the symmetry class
AII of the Hamiltonian $H_0$. The F and B sectors of the sigma model are compact
and non-compact, respectively. This is achieved by demanding that the angles
$\theta_b$, $k_b$ are imaginary while all other angles in Eqs.\ (\ref{qf}),
(\ref{qb}) are real. Below we will find that the saddle point describing the
density of states in the hole occurs with real $\theta_b$. This implies a proper
shift of the integration contour for this angle. 

The sigma model of class AII is designed to study transport properties of a
disordered system. This implies that the $Q$ matrix operates, in particular, in
the RA space allowing for averaging the product of retarded and advanced Green
functions. We are interested in the density of states and hence it suffices to
average just the single retarded function. At the same time, superconducting
boundary conditions implemented in the BdG Hamiltonian (\ref{H}) require to
introduce an additional doubling of fields in the Nambu-Gor'kov space. This can
be achieved within the standard AII class sigma model with the $8 \times 8$
supermatrix $Q$ while the role of  NG space is taken over by the RA structure of
$Q$ (for detailed discussion of the transformation from RA to NG representation
see \cite{supplement, KoziiSkvor}).

Thus we can incorporate the superconducting order parameter directly into the
action of the sigma model,
\begin{gather}
 S[Q]
  = \frac{\pi\nu}{8} \!\int\! d^2r \mathop{\mathrm{Str}} \left[
      D (\nabla Q)^2 + 4(i\epsilon \Lambda - \hat\Delta) Q 
    \right] + S_\theta[Q], \notag\\
 \Lambda
  = \tau_z\sigma_z,
 \qquad
 \hat\Delta
  = \Delta(r) (\tau_x \cos \varphi - \tau_y \sin \varphi).
 \label{action}
\end{gather}
Here $\epsilon = E + i G_t\delta(\mathbf{r}-\mathbf{r}_0)/4\pi\nu$ is the sum
of the energy $E$ and the local dwell term  describing the coupling to a tunnel
tip with dimensionless conductance $ G_t \ll 1$ \cite{supplement}.
 
The action \eqref{action} involves the $\mathbb{Z}_2$ topological term
$S_\theta[Q]$ which appears due to massless Dirac nature of underlying electrons
as was discussed above. The topological term involves only the compact part of
the sigma-model manifold, i.e., its F sector $\mathcal{M}_\text{F}$. In the
general version of the sigma model, that is capable of averaging several
retarded and advanced Green functions, the homotopy group is
$\pi_2(\mathcal{M}_\text{F}) = \mathbb{Z}_2$. However, in the minimal model we
are considering, $\mathcal{M}_\text{F}$ has the structure of the product of two
spheres $S^2 \times S^2$ as seen from Eq.\ (\ref{qf}). In this case the homotopy
group is richer, $\mathbb{Z} \times \mathbb{Z}$. Two integer topological
invariants can be introduced counting the degree of covering of the two spheres
by the mapping $Q$ from real space to $\mathcal{M}_\text{F}$. This allows us to
write the topological term explicitly although in a non-invariant form. With the
parameterization (\ref{qf}), it reads (cf.\ Ref.\ \cite{Ostrovsky}):
\begin{multline}
 S_\theta[Q]
  = \frac{i}{4} \int d^2r \Big[
      \sin\theta_f \big( \nabla \theta_f \times \nabla \phi_f \big) \\
      +\sin k_f \big( \nabla k_f \times \nabla \chi_f \big)
    \Big].
 \label{Stop}
\end{multline}

Let us now analyze the minima of the action (\ref{action}) for the setup Fig.\
\ref{fig_setup}. Circular symmetry of the problem fixes the phase equal to the
polar angle, $\phi_{f,b} = \varphi$. The other parameters depend only on the
radial coordinate $r$. Both angles $\theta$ in F and B sectors obey the Usadel
equation 
\begin{multline}
 D \left[
   \frac{\partial^2 \theta}{\partial r^2}
   +\frac{1}{r}\, \frac{\partial \theta}{\partial r}
   -\frac{\sin 2\theta}{2r^2}
 \right] \\ + 2 i E \sin\theta \cos k + 2 \Delta(r) \cos\theta\label{usadel}
  = 0.
\end{multline}
Inside the hole $\Delta = 0$. At low energies we can also neglect the $E$ term
and the equation becomes independent of $k$. The step-like dependence of the
order parameter, Eq.\ (\ref{Delta}), imposes the boundary condition $\theta(R)
= \pi/2$. There are two possible solutions to the Usadel equation with this
boundary condition:
\begin{align}
 \theta_1
  &= 2 \arctan(r/R), \label{theta1}\\
 \theta_2
  &= \pi - 2 \arctan(r/R). \label{theta2}
\end{align}
Saddle point equations also require $k_f = 0$ and hence the angle $\chi_f$ drops
from the matrix $Q$ and from the action. The two remaining angles $k_b$ and
$\chi_b$ are free and can take any constant values.

The spatial profile of $Q$ is thus fixed by the Usadel equation. The solutions
$\theta_{1,2}$ represent two disconnected saddle points in the F sector while in
the B sector only the saddle point $\theta_b = \theta_1$ is reachable. If the
integration contour for $\theta_b$, which runs along the imaginary axis, is
shifted to the point $\theta_2$ a divergence occurs in the $k_b$ integral. Thus
the B sector is reduced to a hyperboloid parameterized by $i k_b > 0$ and $0 <
\chi_b < 2\pi$, see Ref.\ \cite{supplement}. This is exactly the structure of
the sigma model of class BD as we anticipated from symmetry analysis. The
distinction between even (D) and odd (B) versions of this class is related to
the disjoint character of the manifold due to the discrete degree of freedom in
the F sector. Namely, in class D (B) the two parts of the manifold contribute to
the partition function with the same (opposite) sign. In our problem the odd
symmetry class B occurs. To demonstrate it, we compare the value of the action
(\ref{action}) at the two minima in the F sector. These two solutions indeed
contribute with the opposite sign since the corresponding values of the
topological term (\ref{Stop}) differ by exactly $i\pi$.

The density of states is given by the integral
\begin{equation}
 \rho(E,r)
  = \frac{\nu}{8}\mathop{\mathrm{Re}} \int DQ
    \mathop{\mathrm{Str}} [k \Lambda Q(r)]\; e^{-S[Q]}.\label{rho_integral}
\end{equation}
At low energies, this integral is to be calculated over the saddle manifold.
Apart from the two variables $k_b$ and $\chi_b$ and two disconnected points in
the F sector this manifold involves two Grassmann variables in the matrix $U$,
Eq.\ (\ref{UQU}). In order not to spoil the saddle point, these variables must
be constant in space and fulfill the condition $[U, \hat\Delta] = 0$. We
introduce them according to 
\begin{equation}
 U
  = \exp \left[ \frac{1}{2} \begin{pmatrix}
      0 & \eta \sigma_x + \zeta \sigma_y\\
      i\zeta \sigma_x - i\eta \sigma_y & 0
    \end{pmatrix}_\text{FB} \right].
 \label{U_grassman}
\end{equation}
Within this parameterization, we can rewrite the integral \eqref{rho_integral}
in terms of $k_b$, $\chi_b$, $\eta$, $\zeta$. Thus the problem is reduced to the
0D sigma model of class B (the two disjoint points in the F sector contribute
with opposite signs). Explicit calculation of the integral \cite{IvanovSUSY,
supplement} yields the local DOS in the factorized form
\begin{gather}
 \rho(r,E)
  = \nu\, n(r) f(E/\omega_0), \label{rho_result} \displaybreak[3]\\
 n(r)
  = \cos\theta_1(r)
  = \frac{R^2 - r^2}{R^2 + r^2}, \label{n(r)}\\
 f(x)
  = \frac{\gamma}{\pi(x^2 + \gamma^2)} + 1 - \frac{\sin(2\pi x)}{2\pi x}.
    \label{f(x)}
\end{gather}
Here $\gamma = G_t n(r_0)/2\pi \ll 1$ and the low-energy level spacing is given
by
\begin{equation}
 \omega_0^{-1}
  = 2\nu\int d^2r\, \cos\theta_1(r)
  = 2\pi(\log4 - 1) \nu R^2 
 \label{omega0}.
\end{equation}
The spatial profile of DOS, $n(r)$, is fixed by the solution of the Usadel
equation, while the energy dependence is characteristic for the B class and
contains a narrow lorentzian peak at zero energy. This peak is the zero-energy
MBS broadened due to the finite tunneling time. Note that the width $\gamma$ is
position-dependent.

Integrating $\rho(r, E)$ over space yields the global DoS
\begin{equation}
 N(E)
  = \frac{f(E/\omega_0)}{2\omega_0}.
 \label{N_result}
\end{equation}
In the extreme tunneling limit $\gamma\to0$, this function acquires the
contribution $\delta(E)/2$ and coincides with the result \cite{IvanovSUSY} up
to a factor $2$ due to BdG double counting. The Majorana state appears as a
half of a fermionic level.

Spatial and energy dependence of $\rho(r,E)$ factorize at energies much less
than the Thouless energy $E_\text{Th}=D/R^2$. At higher energies fluctuations of
$Q$ are not important and DOS is given just by a single saddle point.
Approximate solution of the Usadel equation (\ref{usadel}) in the limit $E \gg
E_\text{Th}$ yields \cite{supplement}
\begin{equation}
 N(E \gg E_\text{Th})
  = \pi\nu R^2 \left[1 -
      \big( 2 - \sqrt{2} \big) \sqrt{E_\text{Th}/E}
    \right].
 \label{NhighE}
\end{equation}
Local DOS is close to the normal value $\nu$ everywhere except for a narrow
vicinity of the hole boundary. 

\begin{figure}
\includegraphics[width=0.9\columnwidth]{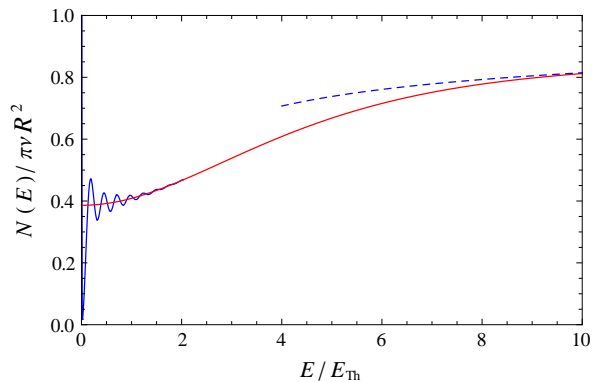}
\caption{(Color online) Global density of states as a function of energy $E$.
The solid blue oscillating curve represents the low-energy result
(\protect\ref{N_result}), the dashed curve shows the high-energy
asymptotics (\protect\ref{NhighE}), and the red curve presents numerical
solution interpolating between the two limits.}
\label{fig_allenergy}
\end{figure}

The local density is measured in the tunneling experiment. The tunneling current
is determined by
\begin{equation}
 I
  = \frac{e G_t}{2\pi\hbar \nu} \int \rho(E,r_0)\big[f(E-eV)-f(E)\big] dE
 \label{I(V)}
\end{equation}
with $f(E)$ being the equilibrium Fermi distribution function. At low
temperatures and voltages, $(T, eV) \ll \omega_0$, we keep only the first
lorentzian term in the energy dependence of the local DOS (\ref{f(x)}). The
differential conductance exhibits a peak
\begin{equation}
 \frac{dI}{dV}
  = \begin{cases}
      \dfrac{e^2\gamma^2}{\pi\hbar\big[\gamma^2 + (eV/\omega_0)^2\big]},
       & T \ll \gamma\omega_0, \\
      \dfrac{e^2\gamma\omega_0}{4\hbar T \cosh^2(eV/2T)},
       & \gamma\omega_0 \ll T \ll \omega_0.         
    \end{cases}
 \label{dIdV}
\end{equation}
At very low temperatures the height of this peak is universal and equals
$e^2/\pi \hbar$. This signifies resonant Andreev reflection at the Majorana
state. This effect in the absence of disorder was studied in Ref.\ \cite{LeeNg}.
At larger (and more realistic) temperatures, $\gamma\omega_0 \ll T \leq
\omega_0$, the height of the peak is parametrically small, $dI/dV \sim
\gamma \omega_0 /T$. 

Noise power of the tunneling current in the same regime $(T, eV) \ll \omega_0$
is \cite{supplement}
\begin{equation}
 S(V,T,r_0)
  = \frac{e^2\gamma\omega_0}{2\hbar},
 \qquad
 \gamma\omega_0 \ll T \ll \omega_0.
 \label{S_VT}
\end{equation}
The noise produced by the Majorana level is $T$- and $V$-independent as long as
$\gamma\omega_0 \ll T$. 

The results (\ref{dIdV}) and (\ref{S_VT}) apply at low temperature and voltage.
When temperature and/or voltage are higher than the level spacing $\omega_0$,
non-zero-energy states also contribute to the tunneling current. Positions and
widths of these states depend on the realization of disorder. For low
temperature but high voltage, $T < \omega_0 < eV$, narrow resonances similar to
Eq.\ (\ref{dIdV}) will occur due to non-zero levels \cite{Paper2}. Positions of
these resonances strongly depend on disorder realization; their heights are
smaller but close to $e^2/\pi \hbar$ and widths are of order $\gamma \omega_0$.
When temperature exceeds $\omega_0$, all the resonances get smeared and the
normal average DOS $\nu$ is recovered.

To conclude, we have studied the local density of states in the superconducting
vortex on the surface of a topological insulator in the superconducting antidot
setup depicted in Fig.\ \ref{fig_setup}. The spatial and energy dependence of
the density of states factorize at low energies and the latter is given by the
0D sigma model of symmetry class B. We have identified the zero-energy Majorana
state occurring due to the topological properties of the system. This Majorana
state exhibits itself via the resonant Andreev reflection at zero energy
yielding the peak in differential conductance with the universal amplitude
$e^2/\pi\hbar$ and width proportional to the normal conductance $G_t$.

We are grateful to B.\ Sacepe for stimulating discussions. This work was
supported by the RFBR grant 10-02-00554-a and by the German Ministry of
Education and Research (BMBF).

\onecolumngrid

\appendix

\section*{Supplementary Information}
\setcounter{page}{1}

\subsection{Derivation of the sigma-model}

\subsubsection{Sigma model for hybrid structure}

In this section we derive the sigma model describing the dynamics of the
topological insulator surface excitations in the presence of a random
potential. We assume the hole geometry depicted in Fig.\ \ref{fig_setup} of the
main text. The derivation starts with the microscopic Bogolyubov -- de Gennes
Hamiltonian (\ref{H}). The sigma model is aimed at averaging the single retarded
Green function determining the density of states. We will also demonstrate the
equivalence of this sigma model to the model of the symplectic symmetry class
AII describing the transport properties of massless Dirac fermions. The latter
model is based on the normal, rather than superconducting, Hamiltonian and is
capable of calculating averaged products of retarded and advanced Green
functions. It is known that the sigma model for Dirac fermions possesses the
specific $\mathbb{Z}_2$ topological term. The equivalence of the two models will
prove the appearance of the topological term in the superconducting case
considered in the present paper.

Let us start with the retarded Green function in the system governed by the
Hamiltonian (\ref{H}). We will use the supersymmetric integral representation
\begin{equation}
 G^R_E(\mathbf{r}, \mathbf{r}')
  = -i \int D\Phi^*\, D\Phi\, \mathop{\mathrm{str}} \left[
      k \Phi(\mathbf{r}) \Phi^\dagger(\mathbf{r}')
    \right] e^{-S},
 \qquad\qquad
 S
  = -i \int d^2r\, \Phi^\dagger \big(
      E + i0 - H
    \big) \Phi.
 \label{SBD}
\end{equation}
Here the vector fields $\Phi$ and $\Phi^\dagger$ contain $4$ commuting and $4$
Grassmann parameters each. Apart from the spin and Bogolyubov -- de Gennes
structure of the Hamiltonian, we also introduce the superstructure in order to
get rid of the normalizing denominator and facilitate further disorder
averaging. The pre-exponential factor contains the supermatrix $k = \{ 1, -1
\}_\text{FB}$ and supertrace is defined as in Ref.\ \cite{Efetov}:
$\mathop{\mathrm{str}} A = A_\text{FF} - A_\text{BB}$.

The superconducting symmetry of the Hamiltonian $H$, Eq.\ (\ref{PH}), gives rise
to specific soft modes -- Cooperons -- of the type $\langle G^R_E G^R_{-E}
\rangle$. These modes are relevant for the average density of states since
they are built out of retarded functions only. In order to include them into
our effective theory, we transform the action by writing half of it in the form
(\ref{SBD}) and another half in the time-reversed (transposed) form:
\begin{multline}
 S
  = -\frac{i}{2} \int d^2r \Big[
      \Phi^\dagger \big(
         E + i0 - H
      \big) \Phi 
      -\Phi^T k \big(
         E + i0 + s_y \tau_y H s_y \tau_y
      \big) \Phi^*
    \Big] \\
  = -\frac{i}{2} \int d^2r
    \begin{pmatrix}
      \Phi^+ \tau_z, & i \Phi^T k s_y \tau_x
    \end{pmatrix} \left[
      (E + i0)\sigma_z \tau_z - H_0 - \tau_z \hat\Delta
    \right] \begin{pmatrix}
      \Phi \\ s_y \tau_y \Phi^*      
    \end{pmatrix}.
\end{multline}
Here we have introduced the matrix $\sigma_z$ operating in the space of
time-reversed blocks -- TR space. The matrix $k$ appeared due to anticommutation
of Grassmann variables. We will denote the doubled vectors as
\begin{equation}
 \Psi
  = \frac{1}{\sqrt{2}} \begin{pmatrix}
      \Phi \\ s_y \tau_y \Phi^*      
    \end{pmatrix},
 \qquad\qquad
 \bar\Psi
  = \frac{1}{\sqrt{2}} \begin{pmatrix}
      \Phi^+ \tau_z, & i \Phi^T k s_y \tau_x
    \end{pmatrix}.
\end{equation}
They are no longer independent, as $\Phi$ and $\Phi^*$ were, but rather obey the
linear constraint
\begin{equation}
 \bar\Psi
  = (C \Psi)^T,
 \qquad\qquad
 C
  = -i s_y \tau_x \begin{pmatrix}
      \sigma_x & 0 \\
      0 & i \sigma_y
    \end{pmatrix}_\text{FB}.
 \label{conjugate}
\end{equation}

At this stage, we are ready to perform the disorder averaging. We adopt the
standard Gaussian white noise disorder model characterized by the correlator
\begin{equation}
 \langle V(\mathbf{r}) V(\mathbf{r}') \rangle
  = \frac{\delta(\mathbf{r} - \mathbf{r}')}{\pi \nu \tau}.
\end{equation}
The single-particle density of states is defined as $\nu = |\mu|/2\pi v_0^2$.
Note that the right-hand side is twice bigger compared to the standard
definition applied to normal metals. The reason for this extra factor $2$ is the
Dirac nature of electron spectrum and hence doubling of the phase space due to
the two types of excitations (electrons and holes). 

Disorder averaging produces the quartic term in the action,
\begin{equation}
 S
  = \int d^2r \left\{
      \frac{(\bar\Psi \Psi)^2}{2\pi\nu\tau} -i \bar\Psi \left[
        (E + i0)\Lambda + \mu - v_0\, \mathbf{s} \cdot \mathbf{p}
        -\tau_z \hat\Delta
      \right] \Psi
    \right\}.
 \label{Spsi}
\end{equation}
Here we use the notation $\Lambda = \sigma_z\tau_z$. The interaction term is
further decoupled with the help of the Hubbard-Stratonovich transformation by
introducing an auxiliary matrix field $Q$. The construction of the nonlinear
sigma model implies that the field $Q$ contains all relevant slow modes of the
disordered system -- diffusons and Cooperons. There are in total three ways to
decouple the four-fermion interaction \cite{Efetov}. One of them involves a
scalar field $\sim \langle\bar\Psi\Psi\rangle$ analogous to the random potential
$V$. This leads only to an irrelevant renormalization of the chemical potential.
Two other decoupling schemes introduce a matrix field with the structure
$\langle\Psi\bar\Psi\rangle$ and $\langle\Psi\Psi^T\rangle$ corresponding to
diffusons and Cooperons, respectively. In order to include all relevant slow
modes in the theory, we have to perform decoupling in both ways, which can be
achieved with a single matrix $Q$. Details of this calculation can be found in
Ref.\ \cite{Efetov}. The resulting action has the form:
\begin{equation}
 S
  = \frac{\pi \nu}{16\tau}\mathop{\mathrm{str}} Q^2
    -i \int d^2r\, \bar\Psi \left[
      E \Lambda + \mu - v_0\, \mathbf{s} \cdot \mathbf{p}
      -\tau_z \hat\Delta + \frac{i Q}{2\tau}
    \right] \Psi.
\end{equation}
The vectors $\Psi$ and $\bar\Psi$ are related by Eq.\ (\ref{conjugate}). This
allows us to limit the matrix $Q$ by the linear constraint $Q = \bar Q \equiv C
Q^T C^T$. This relation keeps only those parameters in $Q$ that couple to the
product $\Psi \bar\Psi$. Integrating out the field $\Psi$ yields the action for
the matrix $Q$:
\begin{equation}
 S
  = \frac{\pi \nu}{16\tau}\mathop{\mathrm{str}} Q^2
    -\frac{1}{2} \mathop{\mathrm{str}} \ln \left[
      E \Lambda + \mu - v_0\, \mathbf{s} \cdot \mathbf{p}
      -\tau_z \hat\Delta + \frac{i Q}{2\tau}
    \right].
 \label{Slog}
\end{equation}

Derivation of the sigma model proceeds with the saddle-point analysis of the
above action. In the dirty limit, the energy and $\hat\Delta$ terms in the
argument of the logarithm are relatively small compared to $Q/\tau$.
With these small terms neglected, the action possesses the uniform saddle point
$Q = \Lambda$. This saddle point corresponds to the self energy of the
average electron Green function in the self-consistent Born approximation.
Other saddle points can be achieved by rotations $Q = T^{-1} \Lambda T$,
if the matrix $T$ commutes with the spin operator $\mathbf{s}$ ($E$ and
$\Delta$ terms are still neglected). Rotations $T$ define the target manifold of
the non-linear sigma model. The effective action of the sigma model within this
manifold is derived with the help of the gradient expansion of Eq.\ (\ref{Slog})
allowing for slow spatial variation of $Q$ and perturbative expansion to the
linear order in $E$ and $\hat\Delta$. Since the $Q$ matrix is trivial in the
spin space, we can safely reduce its size to $8 \times 8$ keeping only Nambu,
TR, and FB structure. Then the self-conjugacy relation acquires the form of
Eq.\ (\ref{barQ}).

The gradient expansion of Eq.\ (\ref{Slog}) is a highly nontrivial procedure in
view of the chiral anomaly of the Dirac fermions. The momentum integrals
arising after the expansion of the logarithm are divergent in the ultraviolet
limit and require a proper regularization. The result of the expansion is
independent of a particular regularization scheme provided the gauge
invariance is preserved. The anomaly affects only the imaginary part of the
action and leads to the appearance of the topological term \cite{Ostrovsky}. At
the same time, the real part can be obtained in a straightforward way since all
the arising momentum integrals are convergent. The result reads
\begin{equation}
 \mathop{\mathrm{Re}} S
  = \frac{\pi \nu}{8} \int d^2r \mathop{\mathrm{str}} \left[
      D (\nabla Q)^2 + 4 i (E \Lambda - \tau_z \hat\Delta) Q
    \right].
 \label{ReS}
\end{equation}
Here the diffusion coefficient is $D = v_0^2 \tau/2$ and the matrix $Q$ is
reduced to the $8 \times 8$ size. The action (\ref{action}), used in the main
text, differs from Eq.\ (\ref{ReS}) only by a $\pi/2$ rotation of the
superconducting phase and by an imaginary contribution to the energy term. The
latter corresponds to a finite dwell time of the electron due to the coupling to
the tunneling microscope tip as we elaborate below.

In order to explain the emergence of the topological term in the imaginary part
of the action, we will prove the equivalence of the sigma model, derived here
for the hybrid structure of Fig.\ \ref{fig_setup}, to the symplectic sigma model
for Dirac fermions derived in Ref.\ \cite{Ostrovsky}.

\subsubsection{Sigma model for Dirac fermions}

The sigma model obtained by the gradient expansion of Eq.\ (\ref{Slog}) belongs
to the symplectic symmetry class AII. This is quite natural since the
Hamiltonian possesses only time-reversal symmetry in the absence of the 
$\hat\Delta$ term. Let us demonstrate the equivalence between our sigma model
(\ref{action}) describing the density of states for the Hamiltonian $H$, Eq.\
(\ref{H}), and the standard symplectic class sigma model describing transport
via the surface states of the topological insulator with Hamiltonian $H_0$. In
the latter case, the sigma model yields averaged products of retarded and
advanced Green functions with the energy difference $\omega$. This requires
introducing retarded-advanced (RA) structure of the action,
\begin{equation}
 S_0
  = -i \int d^2r\, \Phi^\dagger \tau_z \big[
      (\omega/2 + i0) \tau_z - H_0
    \big] \Phi.
\end{equation}
Here the matrix $\tau_z$ operates in the RA space. Subscript $0$ is used to
distinguish this model from the sigma model derived in the previous section and
used in the main text.

The time-reversal symmetry $H_0 = s_y H_0^T s_y$ is taken into account by
further doubling of variables in the TR space,
\begin{multline}
 S_0
  = -\frac{i}{2} \int d^2r \Big[
      \Phi^\dagger \big(
        \omega/2 + i0 - \tau_z H_0
      \big) \Phi 
      -\Phi^T k \big(
         \omega/2 + i0 - \tau_z s_y H_0 s_y
      \big) \Phi^*
    \Big] \\
  = -\frac{i}{2} \int d^2r
    \begin{pmatrix}
      \Phi^\dagger \tau_z, & i \Phi^T k s_y \tau_z
    \end{pmatrix} \big[
      (\omega/2 + i0) \Lambda_0 - H_0
    \big] \begin{pmatrix}
      \Phi \\ i s_y \Phi^*      
    \end{pmatrix}.
\end{multline}
Note that the matrix $\Lambda_0 = \tau_z$ multiplying the frequency term is
different from its counterpart $\Lambda$ from the previous section. The vectors
$\Psi$ and $\bar\Psi$ as well as the charge conjugation matrix $C_0$ are also
defined differently,
\begin{equation}
 \Psi
  = \frac{1}{\sqrt{2}} \begin{pmatrix}
      \Phi \\ i s_y \Phi^*      
    \end{pmatrix},
 \qquad\qquad
 \bar\Psi
  = (C_0 \Psi)^T
  = \frac{1}{\sqrt{2}} \begin{pmatrix}
      \Phi^+ \tau_z, & i \Phi^T k s_y \tau_z
    \end{pmatrix},
 \qquad\qquad
 C_0
  = -i s_y \tau_z \begin{pmatrix}
      \sigma_x & 0 \\
      0 & i\sigma_y
    \end{pmatrix}_\text{FB}.
\end{equation}

Already at this stage we can prove the equivalence of the two models by
selecting a proper unitary rotation of the field vector $\Psi$. This unitary
rotation should obey the following properties:
\begin{equation}
 U^T C_0 U
  = C,
 \qquad\qquad
 U^{-1} \Lambda_0 U
  = \Lambda.
 \label{transform}
\end{equation}
These relations bring the charge conjugation matrix $C_0$ and the matrix
$\Lambda_0$ to the representation used in the previous section for the sigma
model of the hybrid system. There are many matrices $U$ that fulfill identities
(\ref{transform}). One possible choice is
\begin{equation}
 U
  = \begin{pmatrix}
     1 & 0 \\
     0 & i \tau_y
    \end{pmatrix}_\text{TR}.
\end{equation}
In the case of usual rather than Dirac fermions, an analogous equivalence
between the sigma model for the density of states in a superconducting hybrid
structure and the orthogonal class sigma model is discussed in Ref.\
\cite{KoziiSkvor}.

\subsubsection{Topological term}

So far we have discussed the derivation of the real part of the sigma-model
action (\ref{action}) and also proved the equivalence of this model to the
symplectic class sigma model for Dirac fermions (up to the boundary conditions
involving the term $\hat\Delta$). The latter model is known to possess the
$\mathbb{Z}_2$ topological term as an imaginary part of its action, see Ref.\
\cite{Ostrovsky}. This allows us to include the topological term in the action
(\ref{action}) and thus complete its derivation.

The explicit form of the topological term is written in the main text, Eq.\
(\ref{Stop}), in a noninvariant form, using the parameterization (\ref{UQU}) --
(\ref{qb}). In this section we will discuss an indirect but explicitly invariant
representation of this topological term.

The target manifold of the symplectic class sigma model is fixed by the
representation $Q = T^{-1} \Lambda T$ with the constraint $Q = \bar Q$. If we
neglect the Grassmann parameters and consider the central part of $Q$, see Eq.\
(\ref{UQU}), these conditions yield $Q_\text{F} \in O(4) / O(2) \times O(2)$ and
$Q_\text{B} \in Sp(2,2) / Sp(2) \times Sp(2)$. The topological term arises in
the compact F sector of the model. An explicit expression for the topological
term can be written within a construction similar to the Wess-Zumino-Witten
term, cf.\ Ref.\ \cite{Konig}.

We first extend the target manifold by relaxing the condition $Q^2 = 1$. Let us
introduce the matrix $\mathcal{Q} = \bar T \Lambda T$. The only restriction on
this matrix is $\mathcal{Q} = \bar{\mathcal{Q}}$. The F sector of the extended
manifold is $\mathcal{Q}_\text{F} \in O(4)$. It has trivial second homotopy
group, $\pi_2[O(4)] = 0$. This allows us to introduce the third, auxiliary,
coordinate $0 \leq t \leq 1$, such that the real 2D space corresponds to $t =
1$, and continuously extend the matrix field according to
\begin{equation}
 \mathcal{Q}(\mathbf{r}, t)
  = \begin{cases}
      Q(\mathbf{r}), & t = 1, \\
      \Lambda, & t = 0.
    \end{cases}
 \label{ext}
\end{equation}
Such an extension assumes that the physical space has no boundary and can be
viewed as a surface of a 3D ball with $t$ being the radial coordinate. With
this definition of $\mathcal{Q}$, the Wess-Zumino-Witten term has the form
\begin{equation}
 S_\theta[Q]
  = \frac{i \epsilon_{abc}}{24\pi} \int_0^1 dt \int d^2r
    \mathop{\mathrm{str}} \Big[
      \mathcal{Q}^{-1} (\nabla_a\mathcal{Q})
      \mathcal{Q}^{-1} (\nabla_b\mathcal{Q})
      \mathcal{Q}^{-1} (\nabla_c\mathcal{Q})
    \Big].
 \label{WZW}
\end{equation}
Here the indices $a$, $b$, and $c$ take three values corresponding to the three
coordinates $t$, $x$, and $y$, and $\epsilon_{abc}$ is the unit antisymmetric
tensor.

The integrand in this expression (\ref{WZW}) is a total derivative therefore the
result of the integration depends only on the value of $\mathcal{Q}$ at the
boundary of integration domain, i.e., on the physical field $Q = \mathcal{Q}|_{t
= 1}$. This can be checked by an explicit calculation of the variation of
$S_\theta$:
\begin{equation}
 \delta S_\theta[Q]
  = \frac{i \epsilon_{abc}}{8\pi} \int_0^1 dt \int d^2r
    \nabla_a \mathop{\mathrm{str}} \Big[
      \mathcal{Q}^{-1} \delta\mathcal{Q}
      \mathcal{Q}^{-1} (\nabla_b\mathcal{Q})
      \mathcal{Q}^{-1} (\nabla_c\mathcal{Q})
    \Big]
  = \frac{i}{8\pi} \int d^2r \mathop{\mathrm{str}} \Big\{
      Q^{-1} \delta Q \big[ Q^{-1} (\nabla_x Q), Q^{-1} (\nabla_y Q) \big]
    \Big\}.
 \label{deltaS}
\end{equation}
A topological term does not change under continuous variations of the field $Q$
but takes different values in different (disconnected) topological sectors of
the model. The general Wess-Zumino-Witten term has a nonzero variation and
hence does not obey this property. However, the term (\ref{WZW}) is constrained
by $Q^2 = 1$. Under this condition, the variation (\ref{deltaS}) is identically
zero. Hence the Wess-Zumino-Witten term (\ref{WZW}) indeed plays the role of a
topological term. 

The value of the topological term is quantized, yielding the topological charge
of the field configuration, only in a system without boundary. This is not the
case for the geometry considered in the paper. The matrix $Q$ is defined in the
finite region inside the hole (see Fig.\ \ref{fig_setup} of the main text) 
with the boundary conditions fixed by the $\hat\Delta$ term in the Hamiltonian
(\ref{H}). The value of the topological term is not completely fixed in this
case. In particular, the expression (\ref{WZW}) depends not only on physical
values of $Q$ but also on the way it is extended in the third dimension, Eq.\
(\ref{ext}). However, this uncertainty leads only to an uncontrolled imaginary
constant in the action. This constant is the same in both topological sectors of
the model and hence does not alter any observable quantities.

An alternative derivation of the sigma-model action, including the topological
term in the form Eq.\ (\ref{WZW}) is possible within the non-abelian
bosonization formalism. Let us for a moment assume, that the Hamiltonian has an
extra chiral symmetry, $H = - s_z H s_z$. This situation corresponds to the
symmetry class DIII and can be realized, e.g., at the Dirac point in the
presence of a random velocity disorder. The sigma model of the class DIII has
an extended target space, corresponding to the manifold of $\mathcal{Q}$
introduced above. This sigma model possesses the Wess-Zumino-Witten term
(\ref{WZW}), see Ref.\ \cite{Witten}, and is the result of the non-abelian
bosonization of the initial fermionic problem. Non-zero chemical potential
drives the system away from the Dirac point and breaks the chiral symmetry.
This reduces the model to the symplectic symmetry class and restricts the field
by the condition $Q^2 = 1$. The resulting action has the topological term in
the form of the restricted Wess-Zumino-Witten term (\ref{WZW}) as discussed
above.

\subsubsection{Tunneling coupling term}

In this section we discuss the appearance of the imaginary contribution to the
energy due to the presence of the tunneling tip. In order to include the
coupling to an external metallic probe, we extend the Hamiltonian as
\begin{equation}
 \mathcal{H}
  = \begin{pmatrix}
      H & t \\
      t^\dagger & H_M
    \end{pmatrix}.
\end{equation}
Diagonal blocks of this matrix are the Hamiltonian $H$ of the topological
insulator surface (\ref{H}) and the Hamiltonian $H_M$ describing electron
states in the metallic tip. Off-diagonal elements $t$ and $t^\dagger$ are
tunneling amplitudes between the two subsystems. We assume these tunneling
amplitudes are local in space and couple the states at point $r_0$ on the
sample surface to the states at the tip of the metallic probe.

Deriving the sigma model as described in previous sections we arrive at the
expression similar to Eq.\ (\ref{Slog}) but with an extended matrix structure:
\begin{equation}
 S
  = \frac{\pi \nu}{16\tau}\mathop{\mathrm{str}} Q^2
    +\frac{\pi \nu_M}{16\tau_M}\mathop{\mathrm{str}} Q_M^2
    -\frac{1}{2} \mathop{\mathrm{str}} \ln \left[
      E \Lambda - \tau_z \mathcal{H} + \begin{pmatrix}
        i Q/2\tau & 0 \\
        0 & i Q_M/2\tau_M     
      \end{pmatrix}
    \right].
\end{equation}
The quantities $\nu_M$, $\tau_M$, and $Q_M$ refer to the metallic tip. In the
absence of tunneling, $t = 0$, the two subsystems decouple and we obtain
independent sigma models describing topological insulator and the tip. We will
assume the density of states $\nu_M$ to be large, that will fix $Q_M = \Lambda$.
Tunneling coupling is assumed to be weak. This allows us to expand the
logarithm in powers of $t$ and $t^\dagger$. The lowest non-vanishing
contribution appears in the second order of the perturbation theory and yields
the tunneling term in the sigma model,
\begin{equation}
 S_t
  = \frac{1}{2} \mathop{\mathrm{str}} \Big[
      (E \Lambda - \tau_z H + i Q/2\tau)^{-1} \tau_z t
      (E \Lambda - \tau_z H_M + i Q_M/2\tau_M)^{-1} \tau_z t^\dagger
    \Big]
  = -\frac{\pi^2}{8}\, \nu \nu_M \mathop{\mathrm{str}} \big(
      Q \tau_z t Q_M \tau_z t^\dagger
    \big).
\end{equation}
In the last expression we have used the saddle-point relation between $Q$ and
the Green function $(E\Lambda - \tau_z H + iQ/2\tau)^{-1}$ in coincident points.
We introduce normal dimensionless (in units $e^2/h$) tunnel conductance of the
junction $G_t = \pi^2 \nu \nu_M \mathop{\mathrm{tr}} (t t^\dagger)$ (note that
$\nu$ and $\nu_M$ are total rather than spinless densities of states at the
Fermi energy). Assuming $Q_M = \Lambda$, we rewrite the action as 
\begin{equation}
 S_t
  = -\frac{G_t}{8} \mathop{\mathrm{str}} \Lambda Q.
\end{equation}
Thus the tunneling term has the same structure as the energy term and produces
an imaginary contribution to $E$, see Eq.\ (\ref{action}) of the main text.

\subsection{Calculation of the density of states}

\subsubsection{Low energies $E \ll E_\text{Th}$}

The local density of states on the surface of the topological insulator is
given by Eq.\ (\ref{rho_integral}) of the main text. The integral over matrix
$Q$ is to be calculated over the saddle manifold fixed by the solution of the
Usadel equation (\ref{usadel}). There are two distinct solutions
$\theta_{1,2}$, [see Eqs.\ (\ref{theta1}), (\ref{theta2}) of the main text]
for the angle $\theta_\text{F}$. In the F sector of the model, all other
parameters are fixed and thus $\theta_{1,2}$ represent two disjoint parts of the
integration manifold. At the same time, only the solution $\theta_b = \theta_1$
is allowed in the B sector, and two other angles, $k_\text{B}$ and
$\chi_\text{B}$, are free forming a hyperboloid (the parameter $k_\text{B}$
must be imaginary in order to ensure convergence of the integral).

To calculate the density of states from Eq.\ (\ref{rho_integral}), we have to
express $\mathop{\mathrm{Str}}(k\Lambda Q(r))$, $S[Q]$ and $DQ$ in terms of
$k_b$, $\chi_b$, $\eta$, and $\zeta$ for the two disjoint parts of the manifold.
However, a direct use of our parameterization leads to an uncertainty in the
integral over the part corresponding to $\theta_f = \theta_2$. The integral over
$d\eta\, d\zeta$ is zero, while the integral over $k_b$ diverges. We resolve
this
uncertainty by a trick proposed in Ref.\ \cite{IvanovSUSY}. The two disjoint
parts of the integration manifold can be mapped onto each other by the
similarity transformation
\begin{equation}
 Q
  \mapsto T^{-1} Q T,
 \qquad\qquad
 T
  = \begin{pmatrix}
      \sigma_x & 0\\
      0 & 1
    \end{pmatrix}_\text{FB}.
 \label{T}
\end{equation}
We use the parametrization (\ref{qf}), (\ref{qb}) in the domain with $\theta_f
= \theta_1$ and then apply the transformation (\ref{T}) to cover the second
topological sector with $\theta_f = \theta_2$. Owing to the relation (\ref{T}),
integration measure is the same in both sectors. It is given by the
superdeterminant of the metric tensor in the space of $Q$ matrices, see Ref.\
\cite{Efetov}. The integration measure reads
\begin{equation}
 DQ
  = \frac{1}{2\pi}\,\tanh\frac{\kappa}{2}\, d\kappa\, d\chi_b\, d\eta\, d\zeta,
 \label{app_berez}
\end{equation}
with $\kappa = ik_b$ ($k_b$ is imaginary to ensure the convergence of the
integral). The integration runs over the hyperboloid $\kappa > 0$, $0<\chi_b <
2\pi$.

The action in the two topological sectors $\theta_f = \theta_{1,2}$ has the form
\begin{align}
 S_1
  &= -4 i \pi \nu \int d^2r\, \epsilon \cos\theta_1 \left(
       \sinh^2\frac{\kappa}{2} - i \eta \zeta \cosh^2\frac{\kappa}{2}
     \right) - \frac{i\pi}{2}
   = -2 i \pi \tilde x \left(
       \sinh^2\frac{\kappa}{2} - i \eta \zeta \cosh^2\frac{\kappa}{2}
     \right) - \frac{i\pi}{2}, \label{app_s1} \\
 S_2
  &= -4 i \pi \nu \int d^2r\, \epsilon \cos\theta_1\, \cosh^2\frac{\kappa}{2}
     -\frac{i\pi}{2}
   = -2 i \pi \tilde x \cosh^2\frac{\kappa}{2}
     +\frac{i\pi}{2}. \label{app_s2}
\end{align}
Here we have introduced the complex dimensionless energy parameter $\tilde x =
\int d^2r\, \epsilon \cos\theta_1 = x + i\gamma$, with the real part $x =
E/\omega_0$ [see Eq.\ (\ref{omega0}) in the main text] and the imaginary part
$\gamma = G_t n(r_0)/2\pi$. The latter appears due to the coupling to the
tunneling tip. The terms $\pm i\pi/2$ in Eqs.\ (\ref{app_s1}), (\ref{app_s2})
appear due to the topological term $S_\theta[Q]$, see Eq.\ (\ref{Stop}). With
the above expressions for $DQ$ and the action we can calculate the total
partition function of the theory,
\begin{equation}
 \int DQ\, e^{-S[Q]}
  = \int_0^\infty \tanh\frac{\kappa}{2}\, d\kappa\, d\eta\, d\zeta
    \exp\left[
      2 i \pi \tilde x \left(
        \sinh^2\frac{\kappa}{2} - i \eta \zeta \cosh^2\frac{\kappa}{2}
      \right) + \frac{i\pi}{2}
    \right]
  = 1.
\end{equation}
This is exactly what one expects for the partition function of a supersymmetric
theory. Note that the topologically nontrivial sector does not contribute to
this result since the action $S_2$ contains no Grassmann variables.

In order to calculate the density of states, we need the pre-exponential factor
from Eq.\ (\ref{rho_integral}). In our parameterization, it has the following
form in the two sectors:
\begin{equation}
 \frac{\nu}{8} \mathop{\mathrm{Str}} (k\Lambda Q)
  = \nu \cos\theta_1 \times \begin{cases}
      \cosh^2\dfrac{\kappa}{2},
        & \theta_f = \theta_1, \\[8pt]
      \sinh^2\dfrac{\kappa}{2} - i \eta \zeta cosh^2\dfrac{\kappa}{2},
        & \theta_f = \theta_2.
    \end{cases}
\end{equation}

\begin{multline}
 \rho
  = \nu \cos\theta_1 \mathop{\mathrm{Re}}
    \int_0^\infty \tanh\frac{\kappa}{2}\, d\kappa\, d\eta\, d\zeta \Bigg\{
      \cosh^2\frac{\kappa}{2} \exp\left[
        2 i \pi \tilde x \left(
          \sinh^2\frac{\kappa}{2} - i \eta \zeta \cosh^2\frac{\kappa}{2}
        \right) + \frac{i\pi}{2}
      \right] \\
      +\left(
        \sinh^2\frac{\kappa}{2} - i \eta \zeta \cosh^2\frac{\kappa}{2}
      \right) \exp\left[
        2 i \pi \tilde x \cosh^2\frac{\kappa}{2} - \frac{i\pi}{2}
      \right]
    \Bigg\} \\
  = \nu \cos\theta_1 \mathop{\mathrm{Re}}
    \int_0^\infty d\kappa\, \cosh\frac{\kappa}{2} \sinh\frac{\kappa}{2} \Bigg[
      -2 i \pi \tilde x \cosh^2\frac{\kappa}{2} \exp\left(
        2 i \pi \tilde x \sinh^2\frac{\kappa}{2}
      \right)
      +\exp\left(
        2 i \pi \tilde x \cosh^2\frac{\kappa}{2}
      \right)
    \Bigg] \\
  = \nu \cos\theta_1 \left(
      1 - \mathop{\mathrm{Re}} \frac{1 + e^{2 i \pi \tilde x}}{2 i \pi \tilde x}
    \right)
  = \nu \cos\theta_1 \left[
      1 + \frac{\gamma}{\pi(x^2 + \gamma^2)} - \frac{\sin(2 \pi x)}{2 \pi x}
    \right].
 \label{app:rho}
\end{multline}
In the last expression we have used the condition $\gamma \ll 1$. The
result (\ref{app:rho}) coincides with Eqs.\ (\ref{rho_result}) -- (\ref{f(x)})
of the main text. In the tunneling limit $\gamma \to 0$ the lorentzian term
yields a delta function. This implies that the Majorana state is robust in a
closed system and disorder does not smear it.

Note, that for an alternative model of the symmetry class D, when the
topological term is absent, the density of states does not contain the Majorana
delta-peak. Two topological sectors contribute with the same sign yielding
\cite{IvanovSUSY}
\begin{equation}
 \rho_D
  = \nu\cos\theta_1 \left[
      1 + \frac{\sin(2 \pi x)}{2 \pi x}
    \right].
\end{equation}

\subsubsection{High energies $E_\text{Th} \ll E \ll \Delta$}

In the previous section, we have calculated the density of states at low
energies. The result (\ref{app:rho}) exhibits a delta-peak at zero energy and
oscillations which decay at the characteristic energy scale $\omega_0$. These
oscillations appear due to repulsion between low-lying levels with energies $E$
and $-E$. The scale $\omega_0$ is the global mean level spacing inside the hole.
At higher energies $E \gg \omega_0$, the effect of level repulsion can be
neglected and the density of states is given by the mean-field expression
\begin{equation}
 \rho(r,E)
  = \nu\mathop{\mathrm{Re}} \cos\theta(r,E),
 \label{app:qc}
\end{equation}
where $\theta$ is a solution of the Usadel equation (\ref{usadel}) with $k = 0$.
The result (\ref{app:qc}) can be obtained from the sigma model identity
(\ref{rho_integral}) by integrating over $Q$ around the unique global minimum
of the action given by the Usadel equation. Supersymmetry ensures that the
integration with respect to small fluctuations around the minimum yields unity
and hence Eq.\ (\ref{app:qc}) follows.

The Usadel equation is a complicated non-linear equation that cannot be solved
analytically for arbitrary $E$. At relatively low energies $\omega_0 \ll E \ll
E_\text{Th}$, the solution can be found perturbatively in $E$. While linear
correction in $E$ to $\theta$ is purely imaginary and does not change the
density
of states, the second order calculation up to terms $\sim E^2$ is necessary to
obtain the observable result. The spatial profile of this small energy
correction does not factorize as in the main term, see Eq.\ (\ref{rho_result})
in the main text. Second order perturbation theory amounts to solving linear
differential equations, obtained by linearizing Eq.\ (\ref{usadel}), and leads
to very complicated functions. We will not discuss any further details of this
approach here.

\begin{figure}
\includegraphics[width=3.5in]{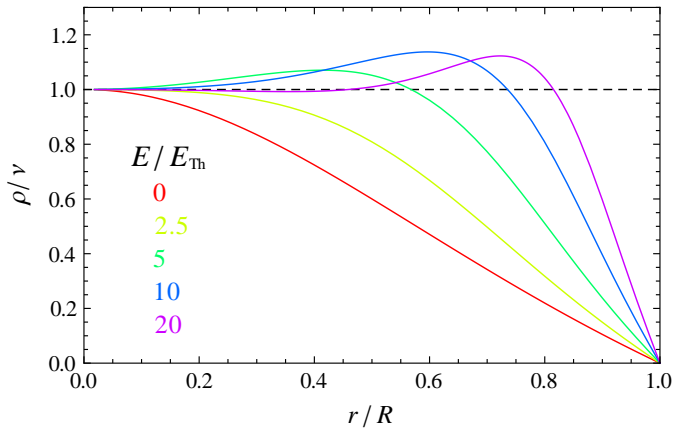}
\caption{Spatial dependence of the density of states for different values of
$E$. At lowest energies, the result (\protect\ref{n(r)}) of the main text is
reproduced, while at energies well above $E_\text{Th}$ the density of states is
depleted only close to the hole edge within the strip $\sim R\sqrt{E_{Th}/E}$ in
agreement with Eq.\ (\protect\ref{thetaHighE}). Remarkably, the density of
states is first increased above its normal value before dropping to zero at $r =
R$.}
\label{fig_spatial}
\end{figure}

As the energy is increased further and exceeds the Thouless energy, $E_\text{Th}
\ll E \ll \Delta$, the Usadel equation admits another approximate solution. The
angle $\theta$ is small everywhere except a narrow vicinity of the hole
boundary. In this vicinity we can neglect the curvature of the superconductor
edge and write an approximate one-dimensional Usadel equation
\begin{equation}
 D\, \frac{\partial^2 \theta}{\partial r^2} + 2 i E \sin\theta
  = 0.
\end{equation}
With the boundary conditions $\theta(R)=\pi/2$ and $\theta(r \ll R) \to 0$,
solution to this equation reads
\begin{equation}
 \theta
  = 4 \arctan \left[
      (\sqrt{2} - 1) \exp \left[
        \sqrt{-2i \frac{E}{E_\text{Th}}} \left(
          \frac{r}{R} - 1
        \right)
      \right]
    \right].
 \label{thetaHighE}
\end{equation}
The density of states is given by Eq.\ (\ref{app:qc}) with the above result for
$\theta$. We see that $\theta$ decreases from $\pi/2$ down to exponentially
small values in a narrow strip of the width $\sim R \sqrt{E_\text{Th}/E}$ near
the hole boundary. The density of states takes its normal value $\nu$
everywhere inside the hole except for this narrow strip where it is depleted
down to zero. Spatial integration yields global density of states
\begin{equation}
 N(E \gg E_\text{Th})
  = \pi\nu R^2 \left[
      1 - (2 - \sqrt{2}) \sqrt{\frac{E_\text{Th}}{E}}
    \right].
 \label{app:NhighE}
\end{equation}
This is the result (\ref{NhighE}) of the main text.

We have also solved the Usadel equation numerically for the whole range of
energies below $\Delta$. The spatial profile of the density of states is shown
in Fig.\ \ref{fig_spatial} and the energy dependence of global DOS is depicted
in Fig.\ \ref{fig_allenergy} in the main text. It perfectly matches both the
low-energy limit (up to delta peak and oscillatory term) of Eq.\
(\ref{rho_result}) and high-energy asymptotics (\ref{app:NhighE}).

\subsection{Tunneling current and noise}

Tunneling current and noise can be found using the Landauer formalism instead
of the sigma model. Electrons incident from the metallic probe are either
reflected normally or experience Andreev reflection. Amplitudes of these
processes can be found from the microscopic Hamiltonian of the system and used
to calculate the current and higher cumulants, such as noise. The current $I(V)$
found this way coincides with the result derived from the sigma model and
presented in the main text, Eq.\ (\ref{I(V)}). The noise, Eq.\ (\ref{S_VT}), was
found by means of the Landauer approach since the sigma-model consideration of
this problem is rather cumbersome. A detailed derivation of the Landauer
formalism for our problem and the calculation of current and noise will be
presented elsewhere \cite{Paper2}.

\end{document}